# Assessing the Prevalence of AI-assisted Cheating in Programming Courses: A Pilot Study

Kaléu Delphino

kdelphino@gatech.edu

*Abstract*—Tools that can generate computer code in response to inputs written in natural language, such as ChatGPT, pose an existential threat to Computer Science education in its current form, since students can now use these tools to solve assignments without much effort. While that risk has already been recognized by scholars, the proportion of the student body that is incurring in this new kind of plagiarism is still an open problem. We conducted a pilot study in a large CS class (n=120) to assess the feasibility of estimating AI plagiarism through anonymous surveys and interviews. More than 25% of the survey respondents admitted to committing AI plagiarism. Conversely, only one student accepted to be interviewed. Given the high levels of misconduct acknowledgment, we conclude that surveys are an effective method for studies on the matter, while interviews should be avoided or designed in a way that can entice participation.

## 1 INTRODUCTION

Generative artificial intelligence (GenAI[1], not to be confused with general artificial intelligence) is the study of models capable of generating rich multidimensional outputs, such as texts, images, audio and video (Leslie & Rossi, 2023). The generation is usually guided by an input text known as the "prompt". For example, giving the prompt "a vase of red flowers" to a GenAI model would generate an image depicting red flowers in a vase.

Practical applications of GenAI are now mainstream thanks to advances in neural networks. In particular, the clever use of attention mechanisms and the subsequent development of the transformer architecture made efficient learning possible over large text corpora (Vaswani et al., 2023). The resulting large language models (LLMs) have incorporated important traits of natural language and are capable of generating human level texts (Piantadosi, 2023). ChatGPT[2], an AI application based on a LLM, can convincingly engage in a conversation and answer questions across multiple subjects (OpenAI, 2022).

---

[1] From now on, we use GenAI and AI interchangeably

Research on applications of LLMs in education is still in its infancy, but looks promising. Personal tutoring systems (Chang, 2022), content explanation (Leinonen et al., 2023) and assignment generation (Jury et al., 2024) are a few of the ideas that have been explored. From another perspective, LLMs are already a reality in schools. ChatGPT and other similar tools are publicly available to students, some free of charge. Even if professors decide to ignore GenAI, students will hardly follow their steps.

A major concern in educational settings is the use of LLMs by students to incur in a novel form of plagiarism: prompting LLMs with assignment descriptions and copying their outputs (Becker et al., 2023; Firat, 2023; Mosaiyebzadeh et al., 2023; Sakib et al., 2023; Sullivan et al., 2023; Zastudil et al., 2023). The risk of academic misconduct through AI is even greater for programming courses (Becker et al., 2023). This is because, somewhat unexpectedly, LLMs have also been shown to be competent in generating *computer code*. Traditionally, programming courses, specially introductory level ones (often referred as CS1), ask students to write small programs in order to teach basic control and data structures (Becker et al., 2023). Now students can simply ask ChatGPT to write those programs for them.

Although the risks of LLMs in Computer Science (CS) education are well recognized, the size of the problem is still unknown. Moreover, students' voices seem to be underrepresented. LLM usage in education is usually approached from the perspective of faculty (Sullivan et al., 2023), and we could only find a few studies trying to elicit students' point of view.

This paper is a first step to fill that gap, by addressing the methodological challenges that arise due to the sensitivity of the matter. We do not expect students to be eager to admit cheating. Therefore, the success of any research project in cheating prevalence assessment depends on the creation of an ethical, trustworthy and inviting environment that fosters honesty.

We conducted a pilot study to assess the feasibility of collecting data from tertiary students through surveys and interviews that can be used for an

---

[2] https://chat.openai.com/

investigation of the prevalence of these new cheating practices, the ways students use AI to cheat, the reasons that make them cheat and their opinions on the fairness of cheating countermeasures, including the ban of AI tools. In particular, we aim to answer the following research questions (RQs):

**RQ1:** Can tertiary students admit cheating in surveys and, if so, are they willing to be interviewed?

**RQ2:** What is the best delivery method to surveys that ask sensitive questions that involve scholar cheating?

Our hope is that an appropriate answer to each of these question may pave the way for future research projects that can answer other RQs such as the following:

**RQ3:** How prevalent is the use of AI-generated code to cheat in tertiary-level programming courses?

**RQ4:** How do students use AI tools to generate code to cheat in tertiary-level programming courses?

**RQ5:** What are the reasons behind the students' decision to cheat with AI tools in tertiary-level programming courses?

**RQ6:** What are tertiary students' views on methods to deal with AI tools in the classroom?

We stress that, as in any pilot study, our main goal here is to answer methodological questions (i.e. RQ1 and RQ2). Nevertheless, we present the results and some initial analysis of the data gathered from subjects of this study in order to show some preliminary directions for answers to RQ3-RQ6.

## 2 RELATED WORK

### 2.1 LLM programming assignment solving capabilities

Following the public availability of LLM-based tools, researchers have been assessing their capabilities to solve CS assignments. They found that AI tools could give correct answers to many assignments (Biderman & Raff, 2022; Denny et al., 2023; Feng et al., 2024; Finnie-Ansley et al., 2022; Hou et al., 2024; Ouh et al., 2023; Puryear & Sprint, 2022; Wermelinger, 2023) and even pass entire courses (Malinka et al., 2023; Savelka et al., 2023). We present a more detailed review in *Appendix 8.1: Review of LLM programming assignment solving capabilities*.

**2.2 Scholar cheating in CS**

Student misconduct has been a problem long before the rise of LLMs in CS. Sheard et al. (2003) reported almost 80% of undergraduate computing students (n≈500) had admitted having cheated. Top reasons to cheat were "not enough time", "will fail otherwise" and "heavy workload". The reported high cheating rate is consistent with the perception that CS is more prone to academic misconduct (plagiarism, in particular), although this can be the result of easier detection (Roberts, 2002).

**2.3 AI-generated text and code detection**

Anti-cheating strategy involves education, prevention, detection and consequence (Sheard et al., 2017). What makes AI-assisted cheating so scary to educators is that it removes detection out of the professor's arsenal. Robust AI-generated text detection is still an open problem. OpenAI, the organization behind GPT, claims it is not possible (OpenAI, n.d.). Researchers are still split between those who believe in the possibility of detection and who do not (Ghosal et al., 2023). We present a more detailed review on the matter in *Appendix 8.2: Review of AI-generated text detection*.

We could not find any theoretical claims about the possibility of detection of AI-generated code. Nevertheless, we argue that detection is probably even harder when the generated text is computer code, as it lacks many of the nuances found in natural language that could serve as features for a detector.

Furthermore, when the AI-generated code to be detected is part of a scholar assignment, things get even more complicated. Detection is only as good as its strength as evidence for academic misconduct proceedings. To allow for a fair process, it needs to be accurate beyond reasonable doubt. In other words, AI code detection needs not only to be fairly good, but at least as good as classical plagiarism detection where original and plagiarized codes can be compared and clearly prove a misconduct had actually happened. As there is no publicly available "original source" when a student uses AI tools, and in light of the previously discussed challenges in detecting AI-generated texts, we are pessimistic about the effectiveness of any detection scheme to deter AI scholar cheating.

## 3 METHODOLOGY

We built and sent a survey to all the students of a cohort of the Education Technology course that is part of the Online Master of Science in Computer Science program at Georgia Tech. That class had about 120 students, which we consider all being part of our population of interest, making our sampling method a census (Callegaro et al., 2015). Additionally, we invited students who had admitted cheating to a short interview.

Regarding data analysis, we used a mixed methods study (Shorten & Smith, 2017). Data gathered from surveys were analyzed in a quantitative way. In particular, we were interested in the proportion of the respondents who had answered in one way or the other to our questions. In case of interviews, we intended to extract some insights from students that cannot be adequately captured in a survey.

The following sections provide more details about the survey and the interview that were applied in the study.

### 3.1 Survey

All the questions of the survey can be seen in *Appendix 8.3: Survey questions*. As the survey contained sensitive questions, we followed advice from survey building literature to increase response rate (Tourangeau & Yan, 2007). More specifically, we have employed the following techniques:

- The survey was self-administered.
- The survey was short (less than 15 questions) and could be answered by clicks/taps only (i.e. no free text fields were present).
- There was an introductory text reassuring respondents that the survey was anonymous and that responses would be disclosed only in aggregate.
- We asked whether the respondent knows someone that has cheated, to obtain indirect data on cheating prevalence and decrease their discomfort when faced with the question of whether they have cheated.
- For the most sensitive question that asks whether the respondent has ever cheated with AI tools, we instead asked how frequently they have done the act, thus avoiding a yes or no answer. We also positioned that question in the middle of the survey to avoid any surprising effects and after gaining some rapport.

The survey was deployed on the Georgia Tech's Qualtrics platform. Students were invited to respond to the survey first by a post on the class online forum. On the next day, an email was sent to all the students. A reminder email was sent a few days after the first email. After a few days without any new response, the survey was closed. The time length between the first and the last response was about 1 month.

## 3.2 Interview

If the survey respondent had admitted cheating, the last question asked if they were available for a short, voice-only interview. Respondents who wished to be interviewed were instructed to send an email to the research team to schedule an appropriate time. We did not directly ask for the respondent's email address in the survey as we wanted to keep it anonymous.

As the number of the consenting interviewees were low, we made a final recruitment attempt through a post on the class forum. Only one interview was eventually conducted. It was made and recorded using Zoom. The interview script can be seen in *Appendix 8.4: Interview script*. All the recruitment posts and emails can be viewed in *Appendix 8.5: Recruitment forum posts and emails*.

## 4 RESULTS

## 4.1 Survey results

A total of 66 responses were obtained in the survey. 4 responses had no answers, so we considered them as nonresponses. Considering our sample had approximately 120 students, we estimate our Response Rate as being (66-4)/120 = 51.67%. Figure 1 shows the number of respondents on each day that the survey was available.

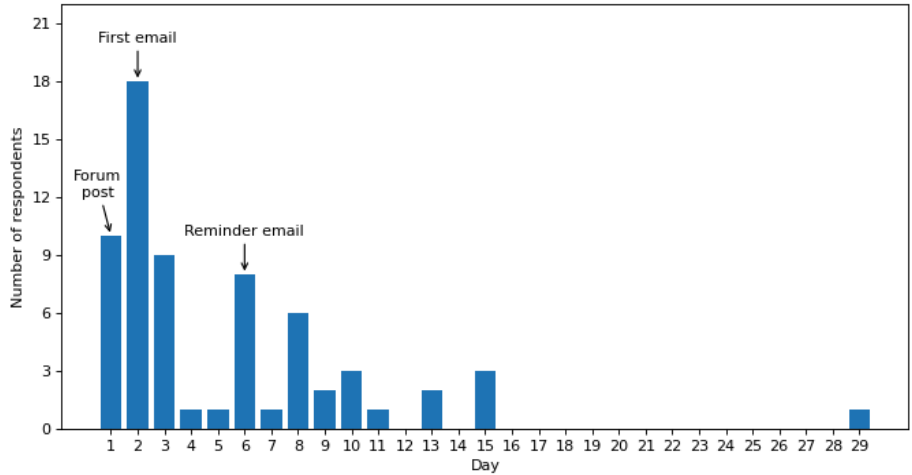

*Figure 1*—Respondents per day

Of the 66 responses, 15 were incomplete and excluded from our analysis. In the remaining 51 responses, 2 of them answered "No" to the first question that asked whether the respondent was a current or former student of the OMSCS program. As our population of interest includes only OMSCS students, these 2 responses were also excluded from our analysis. In the end, 49 responses were considered valid for this study.

Figures 2, 3 and 4 show the results for questions made to assess the prevalence of AI plagiarism in programming assignments. 22% of the respondents reported knowing another student that committed AI plagiarism, while 52% do not definitely discard the possibility of committing it in the future.

As for whether they have already committed AI plagiarism themselves, it is important to highlight that the question was shown only to those who had already answered positively to the previous question that asked whether they have ever generated code using AI tools, a total of 40 respondents. 65% of them answered they have never used AI tools in programming classes when this was not allowed. This means 35% of that respondent subset have admitted using AI tools at least once. When considering all the respondents of the survey, the proportion of students who have admitted AI plagiarism is 28%.

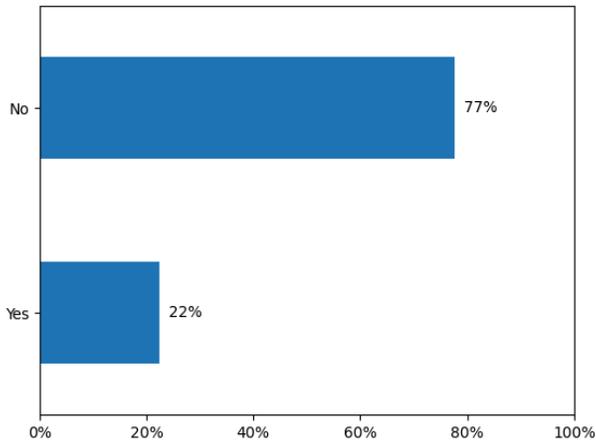

*Figure 2*—Frequency of answers to the question "Do you know another student who has ever used an AI tool to generate code for a programming assignment, when it was not allowed by the professor?" (n=49)

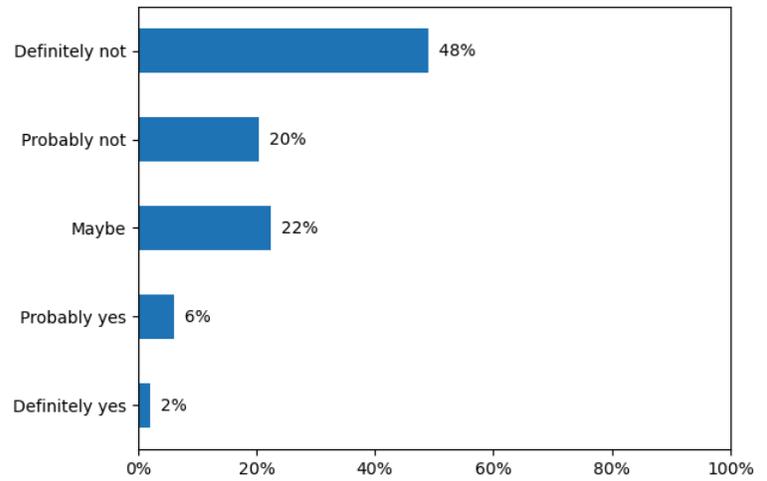

*Figure 3*—Frequency of answers to the question "Do you consider using AI tools to generate code in future assignments, even when it is not allowed by the professor?" (n=49)

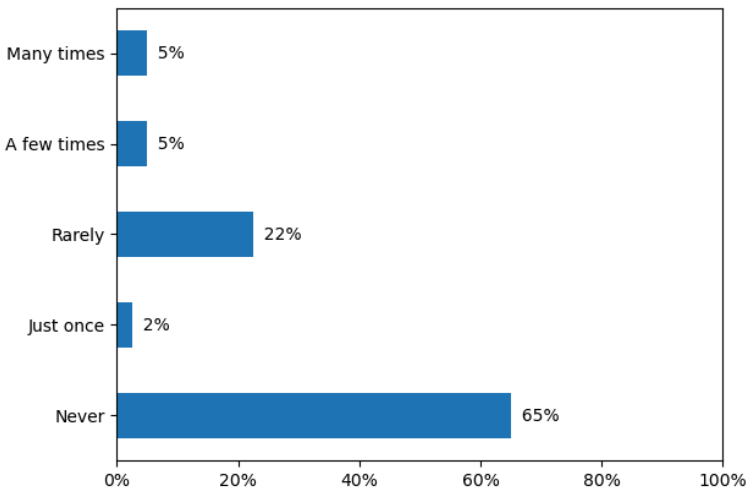

(a)

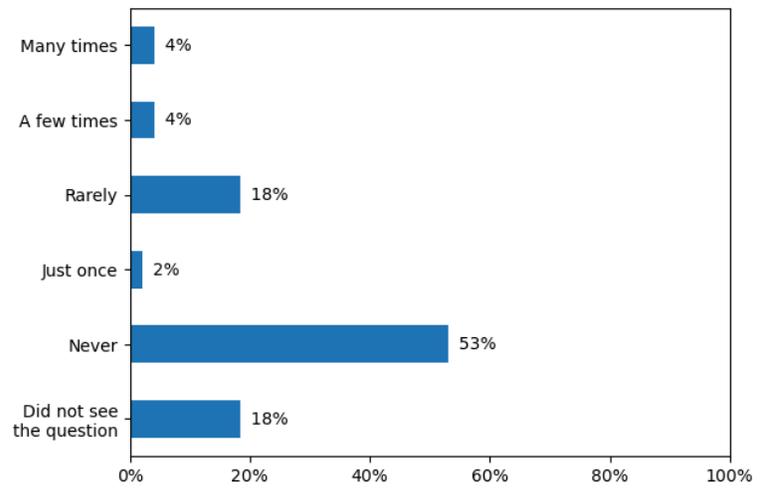

(b)

*Figure 4*—Frequency of answers to the question "How many times have you used AI tools to generate code for programming assignments, when it was not allowed by the professor?"

considering (a) only the respondents who saw the question (n=40)
and (b) all the survey respondents (n=49)

Figure 5 shows the results for the question about how comfortable it was to take the survey. That was the second to last question (only before the interview invitation), after respondents have already been asked the questions about AI plagiarism. 66% of the respondents reported neutral to positive levels of comfort.

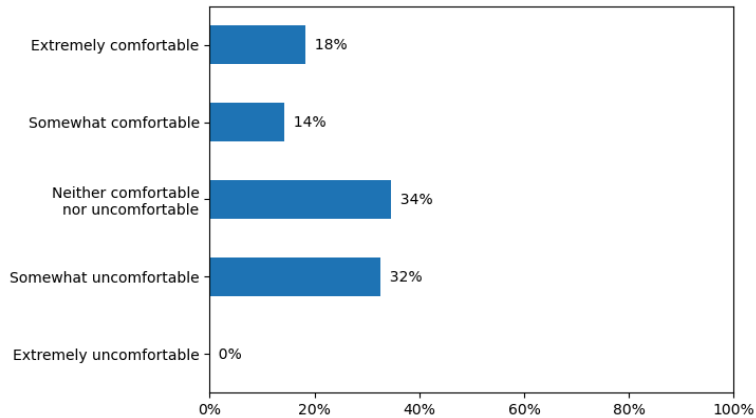

*Figure 5*—Frequency of answers to the question "How comfortable did you feel answering this survey's questions?"

The summaries of the answers for all the other questions can be seen in *Appendix 8.6: Survey results*.

**4.2 Interview results**

Only three people answered positively when asked in the survey if they wished to be interviewed, and only a single one eventually sent the necessary email to be interviewed. The interview ran smoothly and in the expected timeframe of 15 minutes. All questions were answered.

In general, the interviewee reported they had a good experience with ChatGPT, with positive impacts on learning, although they recognize there are still limitations to the tool:

> *"It's really good, it's not like you can just give it a full project and say code this out for me. You sort of have to do it in chunks. So not so much for bigger projects, but for like snippets of code it's OK"*

> *"I think it gives me a good first step in towards understanding, like a good overview and I like that I can ask it questions and it can pinpoint to me very fast what I should focus on. I feel like it has been sort of like a good, like teaching assistant."*

> *"I feel like with stack overflow you probably get two or three options sometimes and you get to see what people like. With ChatGPT it might just give you like one."*

About the way that ChatGPT was used in assignments, they answered:

> *"[To obtain] the structure of the code they wanted, I would put it into ChatGPT and I would give it what I had.". Essentially for it to pass Gradescope because even if the [results] looked similar to what was being asked for, they needed a specific structure to it."*

About the reasons that made they use ChatGPT, having the tool readily available and the perceived benefits were reported as the main drivers:

> *"[T]he availability of it was probably what made me gravitate towards it. And also its way of sort of simplifying some of the what I would call the busy work of coding versus the actual like maybe programmatic knowledge of coding."*

As an advice to educators, the interviewee reported the need of a transparent policy that must be upfront presented to students.

> *"I didn't [disclose AI tool usage], mostly because there was no policy. I used it, but also I didn't say that I used it."*

> *"I think they should be clear on what the policy is when it comes to using ChatGPT and if it's like a no-no, probably share why. And if they allow it, they need to say to what extent and if there are any guidelines for any of that usage of AI […] and I think that should be up in forefront on the course syllabus […] and something that is discussed even in the assignment [instructions]."*

Finally, advice to students include avoiding overreliance:

> *"[V]alidate that what it's giving you is actually true because it has been known to not always be correct."*

> *"[…] I would encourage its use. Personally, I think it's something that is, you know, up and coming and it's sort of inevitable."*

## 5 DISCUSSION

The first research question in this study was "Can tertiary students admit cheating in surveys and, if so, are they willing to be interviewed?". The evidence gathered in this work indicates the answer is yes. About a third of people with previous experience in generating code with AI tools admitted AI plagiarism. This finding matches earlier studies showing high levels of admissions of a variety of socially undesirable behaviors in surveys (Tourangeau & Yan, 2007). We therefore conclude that students are not shy of admitting this specific academic misconduct in anonymous surveys.

Conversely, students were much less keen to be interviewed about the subject. Of the 15 students who were given the opportunity to be interviewed, only one accepted it. In the light of this evidence, we recommend researchers to pay special attention to their interview recruitment methods to make them more attractive, or even avoid interviews on the matter altogether.

Our second research question was "What is the best delivery method to surveys that ask sensitive questions that involve scholar cheating?". In the two delivery methods we used (forum post and email), sending an email directly to respondents proved to be more effective in terms of obtaining a greater number of responses. We do not claim that is the best method, though. We could not test other methods, such as paper surveys.

As for the potential research questions that could be addressed by the methods tested in this study (RQ3 to RQ6), we provide a brief discussion on what the data we gathered indicates to be the answer, although we strongly stress that the surveyed population should not be taken as representative of CS students in general. Further studies, with more representative samples, must be taken to confirm any tendencies we report here.

RQ3 asks "How prevalent is the use of AI-generated code to cheat in tertiary-level programming courses?". The survey responses indicate that prevalence is high, with more than one quarter of the surveyed students admitting it. This results matches earlier assessments of classical forms of plagiarism in CS that were found to be about 20% to 30% (Dick et al., 2003).

RQ4 asks "How do students use AI tools to generate code to cheat in tertiary-level programming courses?". We see from the survey responses that a considerable majority of them use AI tools to fix bugs in their code. This is also the usage mode reported in the interview. One hypothesis is that students prefer to make the bulk of the code themselves and resort to AI tools to help them in getting specific details right.

RQ5 asks "What are the reasons behind the students' decision to cheat with AI tools in tertiary-level programming courses?". Our single interviewee claimed they found the tools handy to deal with "busy work". This corroborates previous studies that cite time pressure as a main driver in cheating behavior (Sheard et al., 2003).

RQ6 asks "What are tertiary students' views on methods to deal with AI tools in the classroom?". The results for the questions on the fairness of some measures show that students oppose the ban of AI tools, favoring adaptive actions, such as more creative assignments. This adaptation, though, should be "soft": weighing exam scores more heavily or bringing paper exams back were not regarded as fair options. A summary of these results can be viewed in *Appendix 8.7: Summary of fairness questions' results*.

**6 CONCLUSION**

The current AI tool offerings and the lack of defenses against them has turned AI plagiarism into a reality in the educational realm. CS education is especially vulnerable due to its reliance on the writing of small portions of code, a task that AI tools have been shown to excel.

Any cost-benefit analysis of potential solutions to this problem needs an estimate of its size. This pilot study demonstrated the feasibility of assessing AI cheating prevalence through surveys, but not through interviews. In the near future, we expect to see surveys being deployed to representative samples of students that would allow us to draw more precise estimates about their AI usage. We also hope that listening to students' voices can help in unalienating them from the important transformations that are now necessary to classrooms.

# 7 REFERENCES

Becker, B. A., Denny, P., Finnie-Ansley, J., Luxton-Reilly, A., Prather, J., & Santos, E. A. (2023). Programming Is Hard - Or at Least It Used to Be: Educational Opportunities and Challenges of AI Code Generation. *Proceedings of the 54th ACM Technical Symposium on Computer Science Education V. 1*, 500–506. https://doi.org/10.1145/3545945.3569759

Biderman, S., & Raff, E. (2022). Fooling MOSS Detection with Pretrained Language Models. *Proceedings of the 31st ACM International Conference on Information & Knowledge Management*, 2933–2943. https://doi.org/10.1145/3511808.3557079

Callegaro, M., Manfreda, K. L., & Vehovar, V. (2015). *Web Survey Methodology*. https://study.sagepub.com/web-survey-methodology

Chang, Y. (2022). Arithmagical: Magical Platform Technology Towards a Strong AI Virtual Tutor. *ACM SIGGRAPH 2022 Appy Hour*, 1–2. https://doi.org/10.1145/3532723.3535464

Cipriano, B. P., & Alves, P. (2023). GPT-3 vs Object Oriented Programming Assignments: An Experience Report. *Proceedings of the 2023 Conference on Innovation and Technology in Computer Science Education V. 1*, 61–67. https://doi.org/10.1145/3587102.3588814

Conroy, G. (2023). Scientific sleuths spot dishonest ChatGPT use in papers. *Nature*. https://doi.org/10.1038/d41586-023-02477-w

Denny, P., Kumar, V., & Giacaman, N. (2023). Conversing with Copilot: Exploring Prompt Engineering for Solving CS1 Problems Using Natural

Language. *Proceedings of the 54th ACM Technical Symposium on Computer Science Education V. 1*, 1136–1142. https://doi.org/10.1145/3545945.3569823

Dick, M., Sheard, J., Bareiss, C., Carter, J., Joyce, D., Harding, T., & Laxer, C. (2003). Addressing student cheating: Definitions and solutions. *ACM SIGCSE Bulletin*, *35*(2), 172–184. https://doi.org/10.1145/782941.783000

Feng, T. H., Denny, P., Wuensche, B., Luxton-Reilly, A., & Hooper, S. (2024). More Than Meets the AI: Evaluating the performance of GPT-4 on Computer Graphics assessment questions. *Proceedings of the 26th Australasian Computing Education Conference*, 182–191. https://doi.org/10.1145/3636243.3636263

Finnie-Ansley, J., Denny, P., Becker, B. A., Luxton-Reilly, A., & Prather, J. (2022). The Robots Are Coming: Exploring the Implications of OpenAI Codex on Introductory Programming. *Proceedings of the 24th Australasian Computing Education Conference*, 10–19. https://doi.org/10.1145/3511861.3511863

Firat, M. (2023). What ChatGPT means for universities: Perceptions of scholars and students. *Journal of Applied Learning & Teaching*, *6*(1). https://doi.org/10.37074/jalt.2023.6.1.22

Gao, C. A., Howard, F. M., Markov, N. S., Dyer, E. C., Ramesh, S., Luo, Y., & Pearson, A. T. (2023). Comparing scientific abstracts generated by ChatGPT to real abstracts with detectors and blinded human reviewers. *Npj Digital Medicine*, *6*(1), 75. https://doi.org/10.1038/s41746-023-00819-6

Ghosal, S. S., Chakraborty, S., Geiping, J., Huang, F., Manocha, D., & Bedi, A. (2023). A Survey on the Possibilities & Impossibilities of AI-generated

Text Detection. *Transactions on Machine Learning Research*.

https://openreview.net/forum?id=AXtFeYjboj

Hou, I., Man, O., Mettille, S., Gutierrez, S., Angelikas, K., & MacNeil, S. (2024). More Robots are Coming: Large Multimodal Models (ChatGPT) can Solve Visually Diverse Images of Parsons Problems. *Proceedings of the 26th Australasian Computing Education Conference*, 29–38. https://doi.org/10.1145/3636243.3636247

Joelving, F. (2023, October 6). Signs of undeclared ChatGPT use in papers mounting. *Retraction Watch*. https://retractionwatch.com/2023/10/06/signs-of-undeclared-chatgpt-use-in-papers-mounting/

Jury, B., Lorusso, A., Leinonen, J., Denny, P., & Luxton-Reilly, A. (2024). Evaluating LLM-generated Worked Examples in an Introductory Programming Course. *Proceedings of the 26th Australasian Computing Education Conference*, 77–86. https://doi.org/10.1145/3636243.3636252

Leinonen, J., Denny, P., MacNeil, S., Sarsa, S., Bernstein, S., Kim, J., Tran, A., & Hellas, A. (2023). Comparing Code Explanations Created by Students and Large Language Models. *Proceedings of the 2023 Conference on Innovation and Technology in Computer Science Education V. 1*, 124–130. https://doi.org/10.1145/3587102.3588785

Leslie, D., & Rossi, F. (2023). *ACM TechBrief: Generative Artificial Intelligence*. Association for Computing Machinery.

Malinka, K., Peresíni, M., Firc, A., Hujnák, O., & Janus, F. (2023). On the
15

Educational Impact of ChatGPT: Is Artificial Intelligence Ready to Obtain a University Degree? *Proceedings of the 2023 Conference on Innovation and Technology in Computer Science Education V. 1*, 47–53. https://doi.org/10.1145/3587102.3588827

Mosaiyebzadeh, F., Pouriyeh, S., Parizi, R., Dehbozorgi, N., Dorodchi, M., & Macêdo Batista, D. (2023). Exploring the Role of ChatGPT in Education: Applications and Challenges. *The 24th Annual Conference on Information Technology Education*, 84–89. https://doi.org/10.1145/3585059.3611445

OpenAI. (n.d.). *How can educators respond to students presenting AI-generated content as their own?* Retrieved February 24, 2024, from https://help.openai.com/en/articles/8313351-how-can-educators-respond-to-students-presenting-ai-generated-content-as-their-own

OpenAI. (2022, November 30). *Introducing ChatGPT*. https://openai.com/blog/chatgpt

Ouh, E. L., Gan, B. K. S., Jin Shim, K., & Wlodkowski, S. (2023). ChatGPT, Can You Generate Solutions for my Coding Exercises? An Evaluation on its Effectiveness in an undergraduate Java Programming Course. *Proceedings of the 2023 Conference on Innovation and Technology in Computer Science Education V. 1*, 54–60. https://doi.org/10.1145/3587102.3588794

Philippe. (2023, August 6). *(Historical) Policy on the use of GPT detectors* [Forum post]. Meta Stack Exchange. https://meta.stackexchange.com/q/391626

Piantadosi, S. (2023). Modern language models refute Chomsky's approach to language. *Lingbuzz Preprint, Lingbuzz, 7180*.

https://lingbuzz.net/lingbuzz/007180/v1.pdf

Puryear, B., & Sprint, G. (2022). Github copilot in the classroom: Learning to code with AI assistance. *Journal of Computing Sciences in Colleges*, *38*(1), 37–47.

Reeves, B., Sarsa, S., Prather, J., Denny, P., Becker, B. A., Hellas, A., Kimmel, B., Powell, G., & Leinonen, J. (2023). Evaluating the Performance of Code Generation Models for Solving Parsons Problems With Small Prompt Variations. *Proceedings of the 2023 Conference on Innovation and Technology in Computer Science Education V. 1*, 299–305. https://doi.org/10.1145/3587102.3588805

Roberts, E. (2002). Strategies for promoting academic integrity in CS courses. *32nd Annual Frontiers in Education*, *2*, F3G-14-F3G-19. https://doi.org/10.1109/FIE.2002.1158209

Sadasivan, V. S., Kumar, A., Balasubramanian, S., Wang, W., & Feizi, S. (2023). *Can AI-Generated Text be Reliably Detected?* (arXiv:2303.11156). arXiv. https://doi.org/10.48550/arXiv.2303.11156

Sakib, N., Anik, F. I., & Li, L. (2023). ChatGPT in IT Education Ecosystem: Unraveling Long-Term Impacts on Job Market, Student Learning, and Ethical Practices. *The 24th Annual Conference on Information Technology Education*, 73–78. https://doi.org/10.1145/3585059.3611447

Savelka, J., Agarwal, A., Bogart, C., Song, Y., & Sakr, M. (2023). Can Generative Pre-trained Transformers (GPT) Pass Assessments in Higher Education Programming Courses? *Proceedings of the 2023 Conference on Innovation and Technology in Computer Science Education V. 1*, 117–123.

https://doi.org/10.1145/3587102.3588792

Sheard, J., Carbone, A., & Dick, M. (2003). Determination of factors which impact on IT students' propensity to cheat. *Proceedings of the Fifth Australasian Conference on Computing Education - Volume 20*, 119–126.

Sheard, J., Simon, Butler, M., Falkner, K., Morgan, M., & Weerasinghe, A. (2017). Strategies for Maintaining Academic Integrity in First-Year Computing Courses. *Proceedings of the 2017 ACM Conference on Innovation and Technology in Computer Science Education*, 244–249. https://doi.org/10.1145/3059009.3059064

Shorten, A., & Smith, J. (2017). Mixed methods research: Expanding the evidence base. *Evidence-Based Nursing*, *20*(3), 74–75. https://doi.org/10.1136/eb-2017-102699

Sullivan, M., Kelly, A., & McLaughlan, P. (2023). ChatGPT in higher education: Considerations for academic integrity and student learning. *Journal of Applied Learning & Teaching*, *6*(1). https://doi.org/10.37074/jalt.2023.6.1.17

Tourangeau, R., & Yan, T. (2007). Sensitive questions in surveys. *Psychological Bulletin*, *133*(5), 859–883. https://doi.org/10.1037/0033-2909.133.5.859

Vaswani, A., Shazeer, N., Parmar, N., Uszkoreit, J., Jones, L., Gomez, A. N., Kaiser, L., & Polosukhin, I. (2023). *Attention Is All You Need* (arXiv:1706.03762). arXiv. https://doi.org/10.48550/arXiv.1706.03762

Weber-Wulff, D., Anohina-Naumeca, A., Bjelobaba, S., Foltýnek, T., Guerrero-Dib, J., Popoola, O., Šigut, P., & Waddington, L. (2023). Testing of detection tools for AI-generated text. *International Journal for Educational*

*Integrity*, *19*(1), Article 1. https://doi.org/10.1007/s40979-023-00146-z

Wermelinger, M. (2023). Using GitHub Copilot to Solve Simple Programming

Problems. *Proceedings of the 54th ACM Technical Symposium on Computer

Science Education V. 1*, 172–178. https://doi.org/10.1145/3545945.3569830

Zastudil, C., Rogalska, M., Kapp, C., Vaughn, J., & MacNeil, S. (2023). Generative

AI in Computing Education: Perspectives of Students and Instructors.

*2023 IEEE Frontiers in Education Conference (FIE)*, 1–9.

https://doi.org/10.1109/FIE58773.2023.10343467

# 8 APPENDICES

## 8.1 Review of LLM programming assignment solving capabilities

Savelka et al. (2023) assessed whether Generative Pre-trained Transformer (GPT, the LLM that ChatGPT is based on) could pass programming courses based on multiple-choice tests. It could not pass in most of the courses, struggling in questions involving output formatting, bug-fixing and refactoring.

Ouh et al. (2023) tried to solve Java programming exercises problems with ChatGPT. It could generate correct and partially correct answers in many instances, but exactly how many have not been reported.

Denny et al. (2023) investigated if Copilot was able to complete programming problems. Copilot is an IDE plugin that gives code suggestions based on Codex, a LLM. 47% of the problems were solved by Copilot in the first attempt. Another 32% could be solved by prompt engineering, i.e. carefully crafting input prompts to produce better responses. They found that pseudocode-like prompts were the most effective, and verbose and conceptual prompts, the least.

Reeves et al. (2023) assessed whether Codex could solve Parsons problems. Parsons problems ask for the correct ordering of code blocks. Some variants also ask for correct indentation. Codex correctly solved around 50% of the problems (80% if indentation was not taken into account). They noted the performance was worse than that of students and that direct and literal prompts seem to make Codex perform better.

Biderman & Raff (2022) studied the capabilities of GPT-J (a variant of GPT) to generate programming assignment answers and avoid detection by MOSS (a code plagiarism detector) at the same time. Researchers generated answers to 7 assignments with GPT-J and compared their MOSS scores to classically plagiarized and non-plagiarized answers. GPT-J generated correct solutions (after some minor human post-processing) to 6 assignments, and 5 of them got similar MOSS scores as non-plagiarized code.

Puryear and Sprint (2022) investigated the effects of Copilot on CS1 courses. They used Copilot to solve programming assignments, and their responses' correctness, appropriateness and style were marked in a 0-5 scale. Copilot's scores were high in general, although he got a perfect correctness score only in one assignment. In a subset of the assignments that were part of a data science course, Copilot scored 84%, less than the average student but still a passing grade. Code plagiarism detectors did not flag Copilot's code.

Cipriano and Alves (2023) assessed whether GPT could solve 6 object oriented programming assignments in Portuguese. For all problems, after some back-and-forth, researchers could generate acceptable solutions, but they would still need to have compilation errors fixed. None of the submitted solutions could pass all the required tests, although the scores were good overall. Researchers noted that the model struggled with assignment rules that forbid some constructs in code, for example, not allowing the use of specific libraries.

Hou et al. (2024) assessed the performance of two LLMs with vision capabilities (GPT-4V and Bard) on visual variants of Parsons problems. Researchers screenshot 6 Parsons problems generated in 4 platforms (for a total of 24 images) and fed them into the LLMs, along with a text prompt directing the LLM to solve the problem. 5 attempts were made for each image and LLM. GPT-4V gave the correct answer in 116 of its 120 attempts (96.7%). Bard gave the correct answer in 69.2% of its attempts. They suggested Bard's worse performance was due to image characteristics, such as resolution. They noted that using prompts that explain what Parsons problems are and number the steps the model needed to follow to solve them had made LLMs perform the best.

Feng et al. (2024) tried to solve Computer Graphics assignments with GPT. They were multiple choice (MCQ) or programming questions that could contain formulas or images. Formulas and images were converted to text before being used as inputs to GPT. From a set of 4 assignments (136 questions), GPT could correctly answer 42.1% (53% of the MCQs; 27.1% of the programming questions in a single attempt, 53.6% in 10 attempts). Performance was worse when questions had images (32.2% with vs. 60% without images). GPT got 64.8% of the questions with formulas (but not images) right. They noted that even when GPT provided the wrong answer in programming questions, sometimes it was possible to easily amend it to get the correct answer. They also highlighted that GPT had performed worse in exercises that demanded inductive instead of deductive reasoning.

Malinka et al. (2023) evaluated ChatGPT CS assignment solving capabilities in a variety of formats (open-ended exams, tests, essays and programming assignments, including course projects). They simulated the use of ChatGPT by a student in 3 different modes: (1) the student copies assignments as prompts and pastes ChatGPT's output back as is; (2) the student interprets ChatGPT's outputs and make corrections when necessary; and (3) the student engage in a conversation with ChatGPT to improve the answers. ChatGPT could obtain scores similar or better than students in open-ended exams (mode 1), worse scores in tests (modes 1 and 2), slightly worse scores in programming assignments (mode 2), and worse scores in essays (mode 4). Mode 1 alone was enough for a student to pass some courses.

Finnie-Ansley et al. (2022) showed that Codex can get scores above the average of human students in real tests used in CS1. They also demonstrated that performance of Codex in a classic problem (rainfall) depends on the formulation of the question.

Wermelinger (2023) qualitatively evaluated Copilot. The author could solve 4 of 8 simple programming problems using code suggested by Copilot alone. He also checked the model's suggestions when solving 7 variants of the rainfall problem. It solved 4.

The previous studies show that LLMs can already do a decent job in programming assignments. Even when they do not perfectly solve an exercise, they can lead a student to the correct answer in many instances. Moreover, given the attention the field is receiving, we expect LLM performance to keep improving.

## 8.2 Review of AI-generated text detection

In a survey paper, Ghosal et al. (2023) points to theoretical works that show fundamental limits in AI text detection. More specifically, there is a bound for the performance of any detector depending on how much the AI and human text distributions overlap (Sadasivan et al., 2023). Watermarking can in principle shift the AI distribution away from the human's, but it is still vulnerable to paraphrasing attacks (Sadasivan et al., 2023). On the other hand, they found studies that show that detection can be performed well if multiple texts are obtained from the same generator, a feasible possibility in practical scenarios (e.g. detecting a social media bot by analyzing its whole post history).

Researchers inclined towards the possibility hypothesis have tried to build detectors in four ways: (1) by adding watermarks to LLM outputs that leave room for posterior detection; (2) by comparing suspicious texts to a database of known texts generated by LLMs (retrieval-based detection); (3) by relying on the statistical differences between human and synthesized text (zero-shot detection); and (4) by training a binary classifier (Ghosal et al., 2023).

Nay-sayers respond that all current detection schemes are susceptible to attacks (Ghosal et al., 2023). In particular, they have been shown to be fooled by paraphrasing, which can be done manually by a human or automatically, although it is still unclear if undetected paraphrased texts become of too low quality in the process. Synthetic text can also be diluted in human text, making detection more difficult.

Besides, the need for at least partial control or parameter knowledge of the text generator is a main drawback of many detection schemes. Watermarking is only effective if an attacker is limited to using watermarked text.[3] Retrieve-based detection needs access to model outputs to store them. Some zero-shooting detection methods, such as DetectGPT, need the knowledge of the generator parameters to work. Others use perplexity scores of the generator as features, thus demanding white-box access.

Empirical evidence so far confirms that detectors are indeed doing a poor job. The question-and-answer site Stack Exchange claims a 32% of error rate when testing a GPT detector over answers (Philippe, 2023). After testing 14 AI text detectors, Weber-Wulff et al. (2023) concluded that "detection tools for AI-generated text [are] neither accurate nor reliable". As expected, traditional plagiarism detectors also fail to detect AI generated text or code (Biderman & Raff, 2022; Gao et al., 2023).

---

[3] Ghosal et al. (2023) raises the possibility of government regulation to require watermarking in LLMs.

In fact, some examples of AI-generated text "in the wild" have only been spotted because someone found an artifact specific to the AI tool in use that had been inadvertently leaked. For example, ChatGPT's common sentence starter "As an AI language model, I..." and the button label "Regenerate response" have been found in scientific papers (Conroy, 2023; Joelving, 2023).

## 8.3 Survey questions

Intro text:

"Hello! My name is Kaléu Delphino. I'm a researcher at Georgia Tech.

I'm researching the use of Artificial Intelligence (AI) tools, such as ChatGPT or GitHub Copilot, by students to generate code in scholar assignments, even when not allowed by the professor. This research can help educators around the world to better design their programming classes.

This anonymous survey is part of that research. Our target respondents are current and former students of the Online Master of Science in Computer Science (OMSCS) at Georgia Tech. It's very short (~5 minutes) and click-only.

Keep in mind that:

1. This survey is completely anonymous and you will not be asked for any details that can otherwise be used to identify you. Not even me will know your identity.

2. Your responses will not be disclosed to the public.

3. Your responses will be mixed up with the responses of the other students and any publication will be in aggregated form.

If you agree to take this survey, please click on the Next button."

Q1 Are you a current or former student of the OMSCS program?

○ Yes

○ No

Q2 How old are you?

○ Under 18

○ 18 - 29

○ 30 - 39

○ 40 - 49

○ 50 - 64

○ 65 or older

Q3 What is your gender?

○ Female

○ Male

○ Other

Q4 Which AI tools have you heard of? (mark all that apply, leave it blank if you haven't heard of any)

- [ ] ChatGPT

- [ ] Copilot

- [ ] Other

Q5 Have you ever generated computer code using AI tools?

- ( ) Yes

- ( ) No

Q6 Do you know another student who has ever used an AI tool to generate code for a programming assignment, when it was not allowed by the professor?

- ( ) Yes

- ( ) No

*Display this question if Have you ever generated computer code using AI tools? = Yes*

Q7 How many times have you used AI tools to generate code for programming assignments, when it was not allowed by the professor?

- ( ) Many times

- ( ) A few times

- ( ) Rarely

○ Just once

○ Never

*Display this question if How many times have you used AI tools to generate code for programming assignments, when it was n... != Never and Have you ever generated computer code using AI tools? = Yes*

Q8 What AI tools did you use to generate code for programming assignments? (mark all that apply)

☐ ChatGPT

☐ Copilot

☐ Others

*Display this question if How many times have you used AI tools to generate code for programming assignments, when it was n... != Never and Have you ever generated computer code using AI tools? = Yes*

Q9 How did you use the AI tools to generate code for programming assignments? (mark all that apply)

☐ I used the assignment's description as an input for the AI tool

☐ I wrote part of the code and asked the AI tool to complete it

☐ I asked the AI tool to fix a bug in my code

- [ ] I asked the tool to convert my code from one language to another

- [ ] Other

Q10 How trustworthy do you think AI-generated code is?

- ○ Extremely untrustworthy

- ○ Somewhat untrustworthy

- ○ Neutral

- ○ Somewhat trustworthy

- ○ Extremely trustworthy

Q11 Do you consider using AI tools to generate code in future assignments, even when it is not allowed by the professor?

- ○ Definitely not

- ○ Probably not

- ○ Maybe

- ○ Probably yes

- ○ Definitely yes

Q12 The following is a list of measures that professors have considered in response to the use of new AI tools by students. For each one, rate how fair you think they are from 1 to 5, where 1 means "definitely unfair" and 5 means "definitely fair"

|  | 1 - Definitely unfair | 2 - Somewhat unfair | 3 - Neutral | 4 - Somewhat fair | 5 - Definitely fair |
|---|---|---|---|---|---|
| Weigh exam scores more heavily | ○ | ○ | ○ | ○ | ○ |
| Ban AI tools in class | ○ | ○ | ○ | ○ | ○ |
| Expose students to the capabilities and limitations of AI tools | ○ | ○ | ○ | ○ | ○ |
| Design AI-proof assignments | ○ | ○ | ○ | ○ | ○ |
| Bring back paper exams | ○ | ○ | ○ | ○ | ○ |
| Use oral, video, and image-based assignments | ○ | ○ | ○ | ○ | ○ |
| Propose more creative assignments | ○ | ○ | ○ | ○ | ○ |
| Making sure all students will have access to AI tools equally | ○ | ○ | ○ | ○ | ○ |

Q13 How comfortable did you feel answering this survey's questions?

○ Extremely uncomfortable

○ Somewhat uncomfortable

○ Neither comfortable nor uncomfortable

○ Somewhat comfortable

○ Extremely comfortable

*Display this question if How many times have you used AI tools to generate code for programming assignments, when it was n... != Never and Have you ever generated computer code using AI tools? = Yes*

Q14 Are you available for a 15-minute interview (voice only) to elaborate on how you used the AI tools in programming assignments (this will count as participation in this class, your name will not be disclosed)?

○ No

○ Yes (please send an email to kdelphino@gatech.edu)

**8.4 Interview script**

1. What is your academic background?
2. Do you remember when you first learned you could use AI tools to generate code?
3. What do you think about the quality of the code generated by AI tools?
4. How do you verify whether the code is in fact correct?
5. Do you think AI tools can already replace forums like Stack Overflow?
6. You have been selected for this interview because you answered that you used an AI tool to generate code for an assignment when it wasn't allowed. Could you provide some context on that?
7. Do you remember the specific tools you used? How did you use them?
8. What were the main reasons that made you opt to use an AI tool?
9. Did you take any steps to conceal the usage of the AI tool for the assignment?
10. How do you think your decision to use AI tools has impacted your understanding of the subject matter?
11. Reflecting on your experience, do you believe there were any alternative solutions or resources that could have helped you overcome the challenges you faced with your programming assignments?
12. How do you think educators should handle AI tools in their programming classes?
13. What advice would you give to other students who may be considering using AI tools to complete their programming assignments?

**8.5 Recruitment forum posts and emails**

*8.5.1 Survey recruitment post*

**Title:** "Survey on AI tools usage in programming assignments"

**Body:** "Dear classmates.

I invite you to respond to this very short survey that is part of my project:

[survey link]

Thank you very much!"

*8.5.2 Interview recruitment post*

**Title:** "AI tool users, make your voice heard!"

**Body:** "While we still have not a reply about Peer Feedback, here is an alternative opportunity to get participation points.

I'm conducting a research project on AI usage by students in programming, especially when it's not allowed by professors. It's not meant to be a witch hunt or anything. On the contrary! Since AI output is currently undetectable (and there is reason to believe it will stay that way), the ideia of this project is to show educators the size of the "problem", so they can make more informed decisions about how to better handle AI tools in the classroom instead of outright banning them.

I intend to make this project into a publishable paper, so this is an opportunity to not only get some participation points, but to actually contribute to the field.

So, if you:

1. have already used an AI tool in a programming class when it was not allowed;

2. want to make a difference in the educational landscape; and

3. want to get participation points for this course

I invite you to participate in a 15-minute, voice-only interview. Your identity will be kept confidential.

Just send me an email at kdelphino@gatech.edu.

And in case you are wondering, people have already been interviewed by me for this project, so no, you won't be the first one :-).”

### 8.5.3 Survey recruitment email

**Title:** "Survey on AI tools usage in programming assignments"

**Body:** "Dear classmates.

I invite you to respond to this very short survey that is part of my project:

[survey link]

Thank you very much!"

### 8.5.4 Survey recruitment email (reminder)

**Title:** "[REMINDER] Survey on AI tools usage in programming assignments"

**Body:** "Hi!

This is a reminder for those who have not answered to my very short survey. If you still have not answered, please do! I only need a few more responses.

The link for the survey is [survey link]

Again, thank you very much for your help!"

**8.6 Survey results**

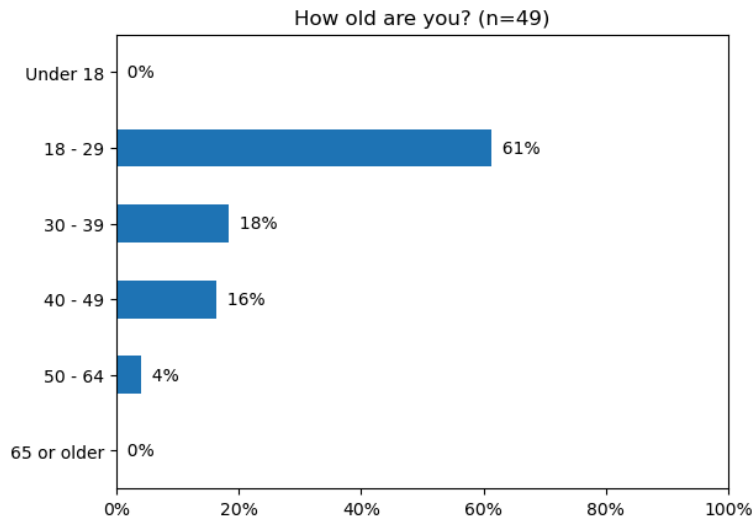

*Figure 6*—Results of Q2

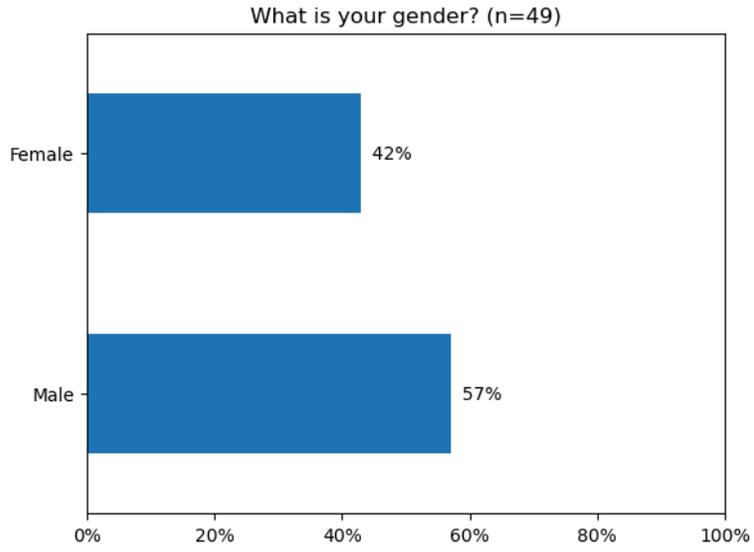

*Figure 7*—Results of Q3

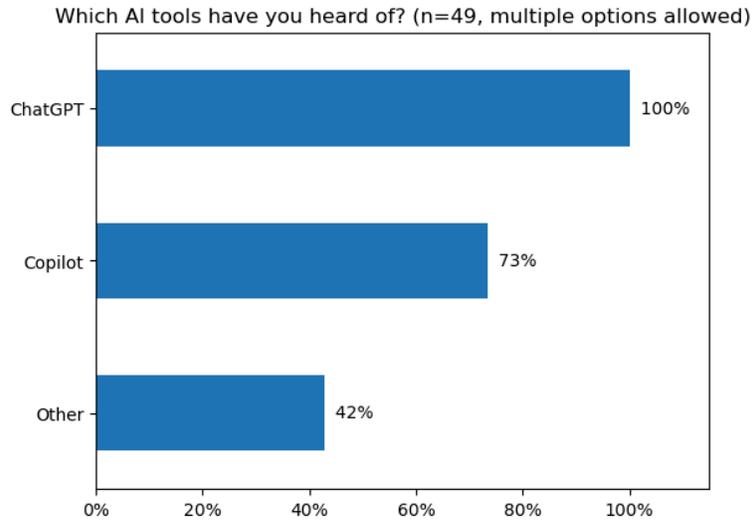

*Figure 8*—Results of Q4

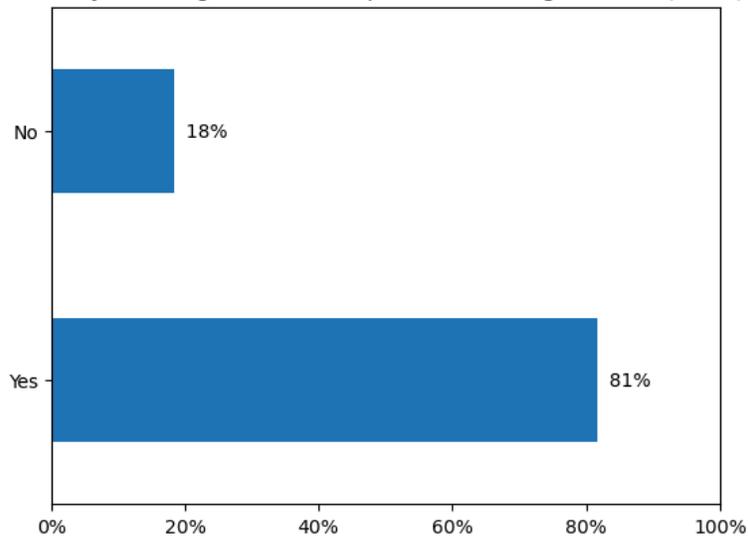

*Figure 9*—Results of Q5

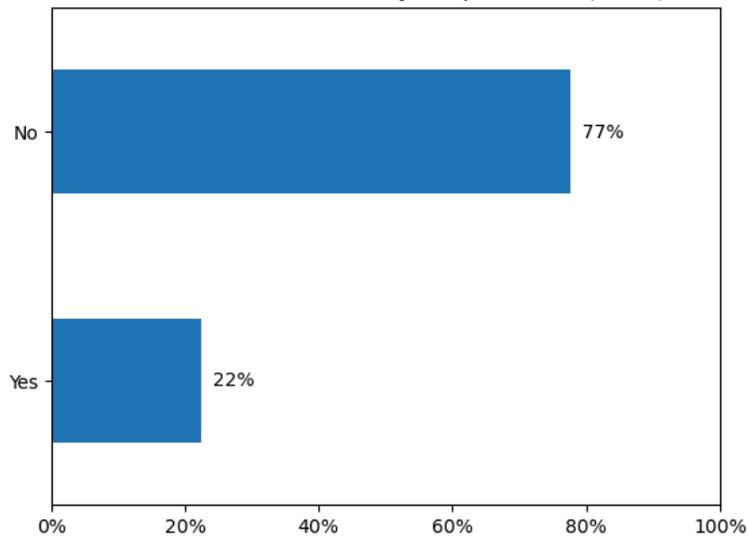

*Figure 10*—Results of Q6

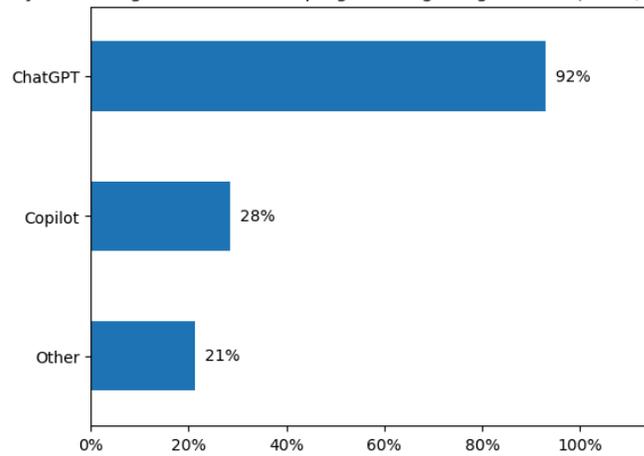

*Figure 11*—Results of Q8

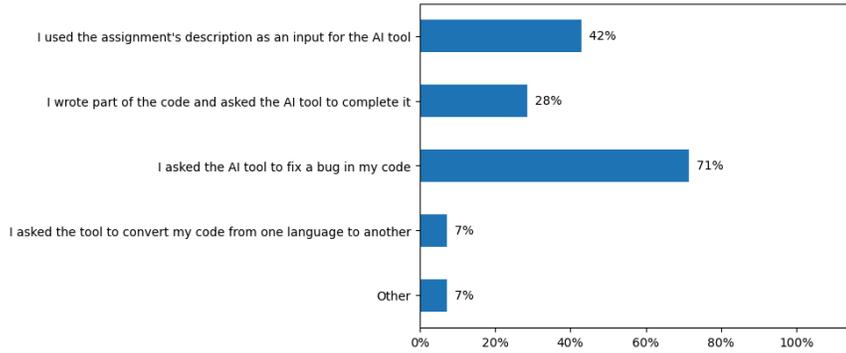

*Figure 12*—Results of Q9

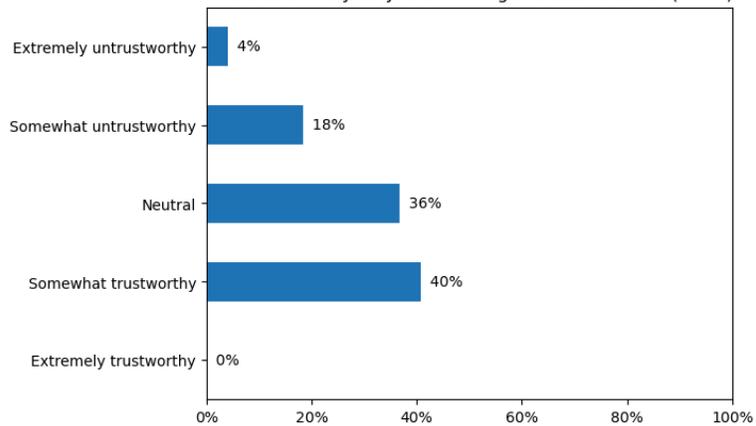

*Figure 13*—Results of Q10

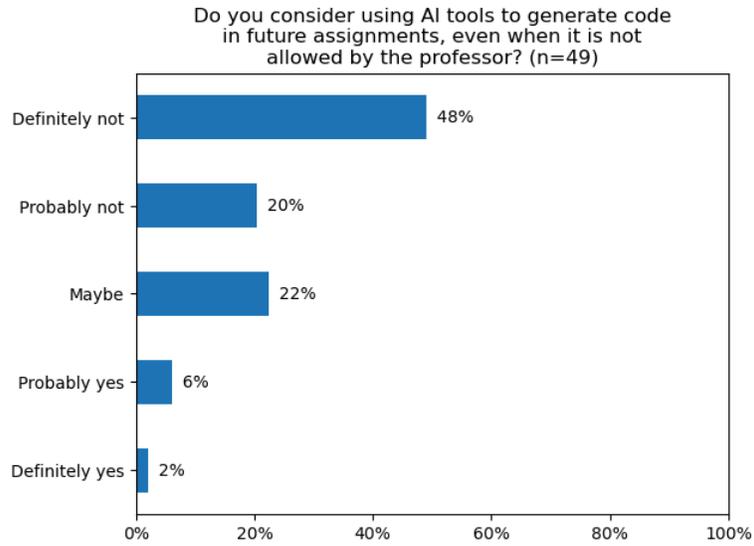

*Figure 14*—Results of Q11

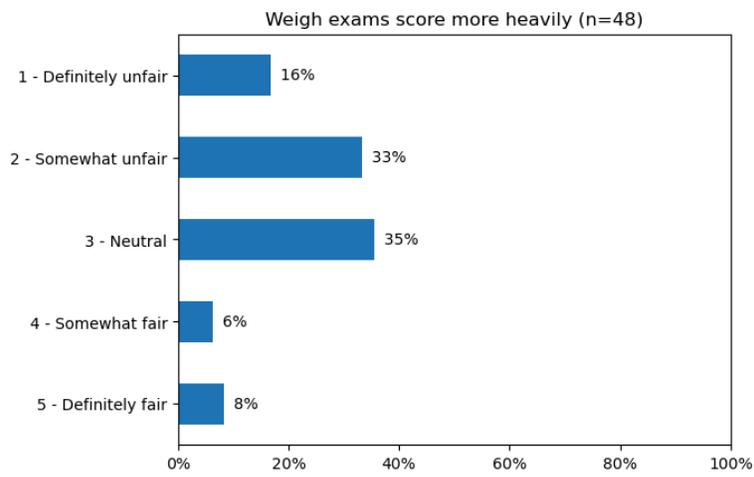

*Figure 15*—Results of Q12-1

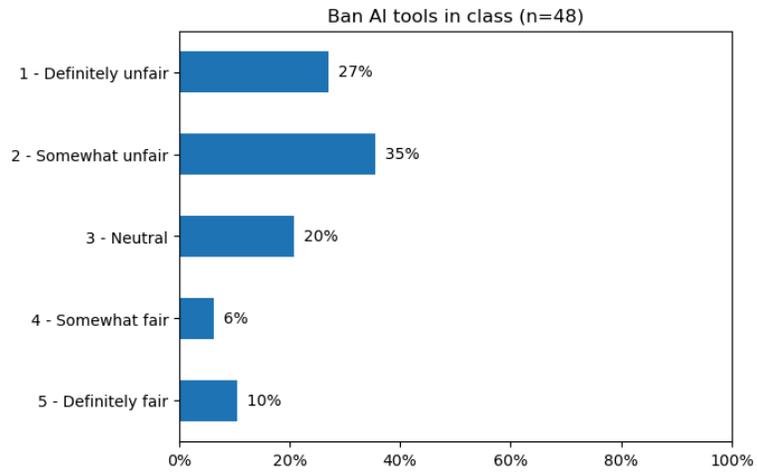

*Figure 16*—Results of Q12-2

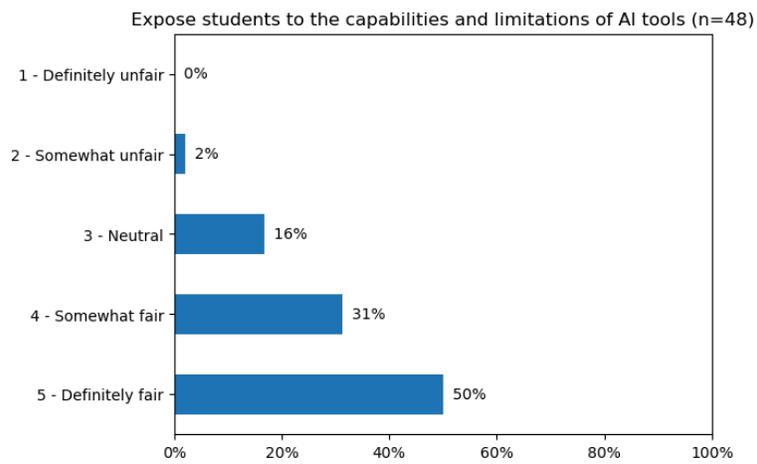

*Figure 17*—Results of Q12-3

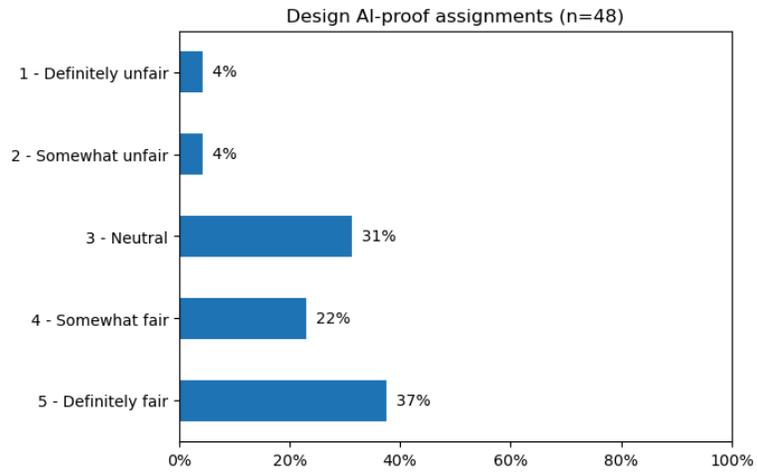

*Figure 18*—Results of Q12-4

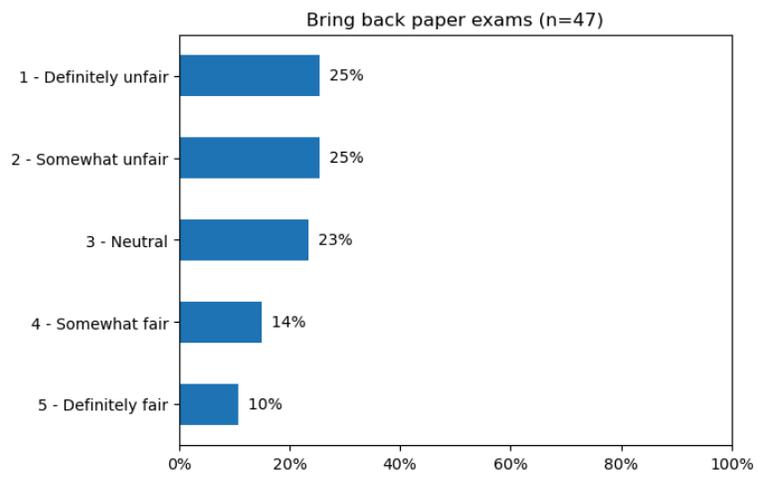

*Figure 19*—Results of Q12-5

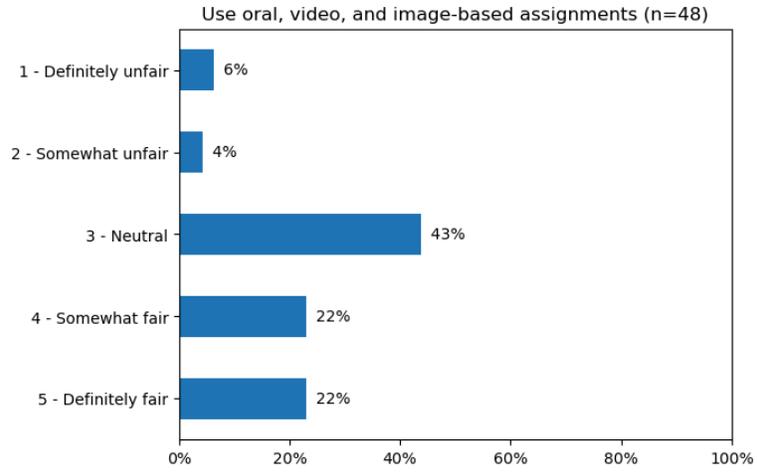

*Figure 20*—Results of Q12-6

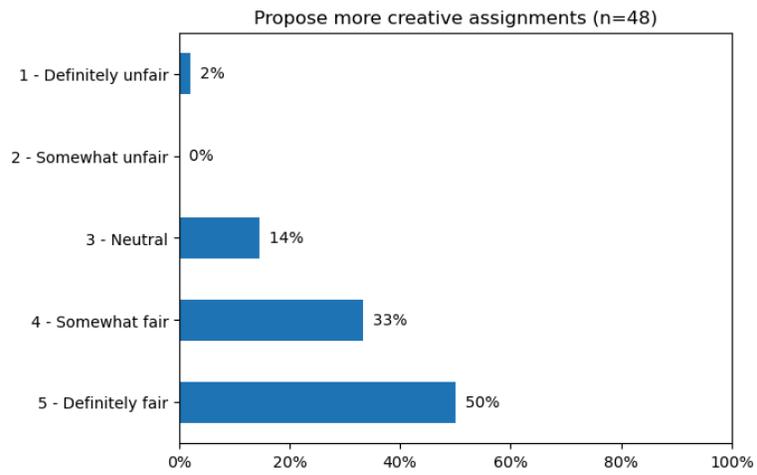

*Figure 21*—Results of Q12-7

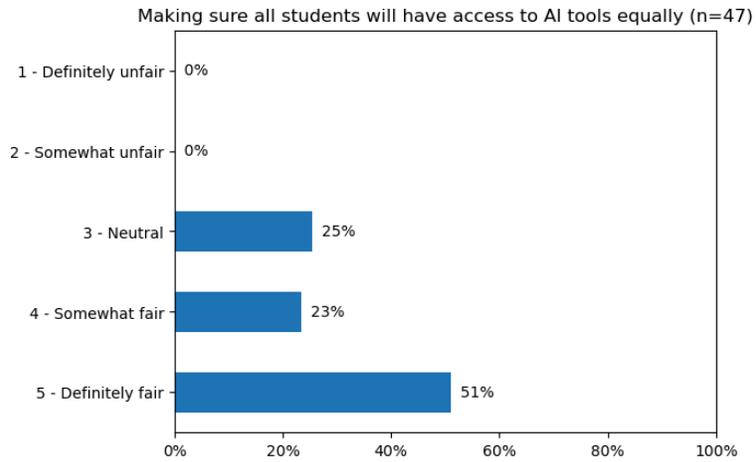

*Figure 22*—Results of Q12-8

## 8.7 Summary of fairness questions' results

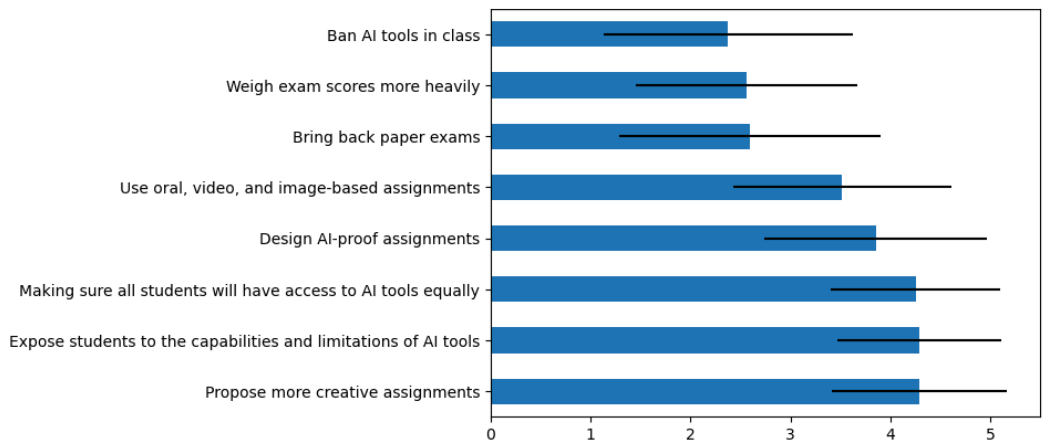

*Figure 23*—Means and standard deviations of the fairness questions